# Malware and Exploits on the Dark Web

Jonah Burgess, *Cyber-Security student, Queen's University Belfast*

*Abstract*— In recent years, the darknet has become the key location for the distribution of malware and exploits. We have seen scenarios where software vulnerabilities have been disclosed by vendors and shortly after, operational exploits are available on darknet forums and marketplaces. Many marketplace vendors offer zero-day exploits that have not yet been discovered or disclosed. This trend has led to security companies offering darknet analysis services to detect new exploits and malware, providing proactive threat intelligence. This paper presents information on the scale of malware distribution, the trends of malware types offered, the methods for discovering new exploits and the effectiveness of darknet analysis in detecting malware at the earliest possible stage.

## I. INTRODUCTION

Whilst the online sale of malware and exploits is not a new phenomenon, it has become more sophisticated in recent years and its increased popularity has had an impact on cyber-security. The existence of marketplaces and forums, where malware and exploits are openly traded, has facilitated the growing demand for hacking-related products and services. Like traditional markets, this has led to further innovation in the development of products and has drawn in expertise from specialists who are tempted by the profitability of the black market.

There are three key factors that have helped crimeware marketplaces; the rise of anonymous crypto-currencies such as Bitcoin and Litecoin, the availability of anonymity networks such as Tor and the increased attack surface triggered by the growing size of the internet and the available platforms such as mobile and cloud computing [3]. Dark Web marketplaces such as The Silk Road, Evolution and AlphaBay have been subjected to growing media attention in recent years which has exposed these sites to new users. Where traditionally these marketplaces may have been difficult to find for those who are not in the know, they have now become accessible for anybody with an interest or curiosity.

Although there have been several high-profile takedowns of Dark Web marketplaces (e.g. the arrest of Ross Ulbricht and seizure of The Silk Road), large scale scamming of customers (e.g. the exit scam performed by the Evolution marketplace) and arrests of malware authors and distributors (e.g. the arrest of the Blackhole Exploit Kit developer/admin), there has been no sign of their popularity declining. When a marketplace is taken down by law enforcement, several more are ready to fight for their position as market leaders. As law enforcement cracks down, marketplace operators and forum administrators put a larger focus on security and anonymity and the community becomes more secretive and suspicious of outsiders.

Products offered on marketplaces and forums are varied in terms of type (e.g. Trojan, Virus, Worm, RAT, Exploit kit), price and required technical ability. There has been a rise in as-a-service offerings such as DDoS-as-a-service, spam-as-a-service and hacking-as-a-service. The availability of these services has enabled criminals with limited technical abilities to achieve their goals, utilizing the abilities offered by vendors and service providers. Zero-day exploits are commonly listed on DarkNet markets and although legitimate bug bounty programs have helped to discourage security researchers from selling zero-days on the black market, the higher profitability (10-100 times more) means that this is still a popular option.

### A. The Dark Web

The Dark Web refers to a collection of websites that exist on an encrypted network and cannot be found on traditional search engines or accessed using standard web browsers. The Tor network is used to access these sites whilst providing anonymity to its users. Operating through a network of relays, Tor uses multiple layers of encryption to ensure that no single node ever has access to both unencrypted data and traffic information that could be used to identify users. Using the Tor browser, users can access hidden services which are identified by their .onion address, indicating that they can only be accessed using a special browser.

### B. Purchasing Malware and Exploits

The Dark Web hosts many hacker forums and marketplaces that offer malware, exploits, DDoS and hacking services etc. These products and services are purchased through DarkNet marketplaces that resemble legitimate electronic marketplaces such as eBay, both in terms of their appearance, structure and functionality. When buyers purchase a product, they are granted the ability to leave feedback which allows sellers to build their reputation based on the satisfaction of their customers. This feature means that the reliability of the vendors can be easily determined and with help from escrow payment systems, it ensures that a minimal number of buyers fall victim to scams in what is an inherently untrusted market area.

Submitted March 2017. This work was supported in part by Queens University, Belfast.

J. Burgess is with the Centre for Secure Information Technology (CSIT), Belfast, UK (e-mail: jburgess03@qub.ac.uk).

The method of payment is typically crypto-currencies such as Bitcoin and the delivery method electronic e.g. email or download link. This provides anonymity between buyers and sellers who never have to meet in person or exchange personally identifiable information. The online payment and delivery method maximises the accessible market as buyers and sellers can operate in different geographical locations. It also maximises the potential profits as there are no physical production or delivery costs. Many vendors will offer demonstrations of their products or even trials, in an effort to provide verification that their listing is accurate. They may also offer guarantees e.g. malware will be undetectable by 80% of AV software for 72 hours.

## II. PRODUCTS

### A. Tools

Initial Access Tools are offered in the form of exploit kits and zero-day vulnerabilities which can be used to deliver payloads and gain persistent access to systems. Viruses, worms and Trojans are offered with unique obfuscation techniques where sellers use packers, crypters, binders and other obfuscation methods to provide buyers with a unique piece of malware that will bypass detection methods. Some vendors will offer a guarantee on the number of hosts that will be infected, the length of time before malware will be detected by a majority of AVs etc. Some examples of exploit kits sold on the DarkNet in the past include Icepack, Nuclear, LuckySploit and CrimePack.

Zero-day exploits can be extremely profitable but more difficult to sell for a couple of reasons; they are expensive and specialised. Achieving a successful trade requires the existence of a buyer who can afford to purchase a zero-day and has a need for the specialised product. Similarly, there needs to be a seller who has the technical abilities or contacts to obtain a zero-day and the reputation to sell it. Zero-days are often used in highly targeted attacks by skilled individuals or groups that make up a small portion of the buyers' market. The true scale of the zero-day market may be unknown as buyers and sellers may often make deals privately or advertise in highly vetted arenas.

### B. Services

The as-a-service offerings provided by hackers on crimeware marketplaces are not new but over the last 5-10 years, the types of services offered have largely shifted from adware, spyware and spam services to DDoS and exploit services. The Blackhole Exploit Kit was the most popular as-a-service exploit kit from 2010 until 2013, when the owner/operator was arrested. 95% of all malicious URLs found by security researchers in the second half of 2011 came from Blackhole and it is estimated that in 2012, more than half of web threats were a result of the use of this service. [2] Exploit services such as this can be leased for a period of time or for a defined number of infections, which can often be tracked by the seller.

DDoS capabilities have drastically increased over the past decade and hackers controlling large botnets may rent these to buyers as a service. The uses for rented botnets include spamming, cracking and denial of service. These services are generally rented on an hourly, daily, weekly or monthly basis depending on customer needs. Sellers often provide guarantees e.g. website guaranteed to be knocked offline for 24 hours. Hacking-as-a-service is becoming more popular and freelance hackers and hacker-groups offer to loan their skills to customers who have a specific target they want to hack but don't have skills to carry it out themselves. These as-a-service products aim to minimise the technical ability required to get involved with various aspects of cyber-crime and their increased popularity is likely to lead to more technically illiterate criminals acquiring them in the future.

## III. TRENDS

### A. Pricing

The pricing of tools and services on DarkNet marketplaces can vary significantly. The pricing depends on the quality of the product/service, the reputation of the seller and the potential return on investment. When a new piece of malware is released or a relatively new exploit is developed for a vulnerability, the prices will start off high but as vendors release patches and more users begin to apply them, the value declines. The black market reacts similarly to legitimate markets so when there is an abundance of similar products/services offered by different vendors, competition drives down prices and increases quality.

Exploit Kits typically range from $15-$10,000 and exploit services vary depending on the length of the license e.g. in 2011, Blackhole v1.2.1 cost $700 for three months or $1500 for a year whilst Robopak cost $150 a week or $500 a month. Zero-day vulnerabilities range from $500-$300,000 depending on the buyer/seller involved, the severity of the vulnerability, sophistication of the exploit, the age of the zero-day and whether the vendor is currently aware of its existence. Examples of zero-day listings from 2012 show price ranges of $60,000-$120,000 for Windows, $100,000-$250,000 for iOS, $30,000-$60,000 for Android and $5,000-30,000 for Adobe Reader [2]. These figures may not accurately represent the true value of the zero-day market since many high value purchases are likely to take place in highly vetted marketplaces and forums with more secrecy focused buyers/sellers.

### B. Sophistication

Trends are showing an increase in the sophistication of the DarkNet marketplaces as more people flock to them and operators/administrators learn from the mistakes of predecessors who have been taken down or arrested by law enforcement. This trend is supported by the increase in sophistication of the underlying technologies such as anonymity networks, crypto-currencies and internet-connected devices e.g. IoT and mobile computing.

The sophistication of the marketplace/forum users is also increasing as younger participants move up the ranks, these users have grown up using technology and are exposed to a greater availability of training and resources. Where there was previously an influx of traditional criminals flocking to the DarkNet in an effort to expand to cyber-crime, there is now a generation of users who have gravitated towards these marketplaces and forums due to early exposure.

There has also been an increase in the sophistication of the products and services offered. This trend largely correlates with cyber-security in general where attackers are forced to continuously innovate in order to keep ahead in the cat-and-mouse game against defenders. The increased popularity and profitability of the crimeware market has created a higher demand for skilled vendors offering sophisticated products and this has in turn led to a greater variety of options for buyers, competition which forces sellers to offer more sophisticated, easy-to-use services to add value and attract different types of users e.g. those with less technical abilities.

*C. Availability*

Trends show an increase in the availability of hacking tools and services. A rise in the number of Dark Web marketplaces/forums and customers interested in using them has attracted more vendors and skilled individuals who work to develop a wider range of products. In 2016, researchers developed an analysis framework for DarkNet markets and forums and they found 1,573 hacking related products from 434 vendors between 27 marketplaces. They also analysed 21 forums and found 4,423 topics and 31,168 hacking related posts from 5391 users [1]. The increasing availability of products on the black market is likely to keep pace with the increased demand.

*D. Zero-Days*

The zero-day market is rising in popularity and this is leading to an increase in related malware creation and attacks. A research study in 2012 indicated that after zero-day disclosure, the number of attacks increases 2 to 100,000 times. Another piece of research used DarkNet analysis techniques over a 4-week period and successfully detected 16 zero-day exploits including an IE 11 remote code execution exploit listing for 20 BTC and an Android WebView exploit for 40 BTC [2]. Initially zero-days have a high value but once they are known to software vendors and a patch can be released, the value decreases significantly. It can also be difficult for sellers to find a suitable buyer due to the specialised nature of zero-days and their prevalence of use in highly targeted attacks.

*E. Resilience*

The resilience of Dark Web marketplaces has been demonstrated by their ability to survive the increasing waves of takedowns. The increased technical skills and level of international cooperation of law enforcement, combined with the ever-growing popularity of marketplaces has led to many closures and arrests. Despite this trend, new DarkNet markets quickly pop up to replace closing markets and they develop new security measures to prevent themselves succumbing to the same fate. The community of sellers and buyers also become more careful with the assistance of comprehensive documentation designed to help users maintain OpSec e.g. guides on how to use PGP encryption and VPNs. Like legitimate markets, the resilience of crimeware marketplaces is evident in their ability to keep pace with ever-changing technology and increasing demand.

## IV. IMPLICATIONS

Measuring the effect that DarkNet markets have had on the overall malware industry and the security of the internet is a complex task. We have already seen evidence that hacking products and services have increased in abundance and sophistication. The research discussed previously regarding the Blackhole Exploit Kit demonstrated the potential spread of malware when a high quality, maintained exploit kit is available for rent on the DarkNet. It is safe to say that as supply and demand increases and more actors get involved in the DarkNet hacking industry, the volume and complexity of cyber threats seen in the wild will increase.

The prevalence of zero-day vulnerabilities is also having a clear impact and whilst we cannot associate all of the recent trends with the DarkNet markets, we can assume that they are a large contributing factor, since they are the primary vessel for the distribution of zero-day exploits. These exploits are often used in highly targeted attacks in what are often called Advanced Persistent Threats (APTs), with the typical zero-day attack lasting between 19 days and 30 months (312 days on average). Because attacks are normally targeted, it can take a long time before they are discovered in the wild.

A study of three exploit archives found that 15% of the exploits were developed before the vulnerability was disclosed, this indicates that attackers are developing or purchasing exploits before the vendor and AV companies are aware of the issue. After disclosure, the number of attacks increases significantly as attackers move quickly to exploit targets before the vendor releases a patch and users have the chance to apply it. One study showed that after disclosure, the number of malware variants exploiting the vulnerability increases by 183-85,000 per day and the number of attacks increases 2-100,000 times [4].

An example of the impact that zero-days on the DarkNet markets can have in real terms was seen with the Dyre Banking Trojan which was designed to steal credit card information. This malware affected 57.3% of organisations around the globe and utilised a zero-day exploit found in Windows. Microsoft disclosed the vulnerability in February 2015 and within one month, a working exploit was available for sale on a DarkNet market for 48 BTC ($10,000-15,000). It was 5 months from this date before a security firm discovered that the Dyre Banking Trojan used this exploit.

## V. DETECTION AND PREVENTION

There are some commercial products available which provide active threat monitoring on the DarkNet such as the solution offered by RepKnight. However, most of the promising work comes from the research field. A notable example of this is seen in [1] where researchers developed a system to collect cyber threat intelligence from the DarkNet. Through the analysis of forums and marketplaces, they were able to detect approximately 305 threats per week. During their research, they detected 16 zero-day exploits from marketplaces over a 4-week period. At the time of publishing, the system was monitoring 27 marketplaces and 21 forums.

The system is composed of two separate crawlers which traverse marketplaces and forums, retrieving only relevant webpages. Two parsers are then applied to extract useful information such as product title, description, vendor name, CVE and ratings from marketplace data. Information mined from forum data includes topic and post content, author, status and reputation. The parsed information is in turn passed to a classifier which uses a machine learning algorithm to detect and label relevant content, allowing irrelevant products/topics to be ignored e.g. drugs, weapons etc.

A set of marketplaces and forums were used for training the system where 25% of the data was labelled manually by security experts. After the manual labelling and training period, system testing showed promising results; 87% of marketplace relevant marketplace data was recalled with a precision of 85%. Relevant forum data was recalled at 92% with a precision of 82%. The system was also used to analyse cross-site connections, showing when vendors listed products on multiple sites and forums. This allowed the researchers to analyse the social grouping of hacker networks and provide useful insights in terms of vendor reliability and popularity.

Another notable example of research is seen in [4] which focuses on zero-day detection. Although this research does not concentrate on DarkNet markets or forums, its findings may assist with the early detection of zero-days which are often traded on the DarkNet. The method for identifying zero-day vulnerabilities involves gathering information from the Worldwide Intelligence Network Environment (WINE) which includes data harvested from 11 million hosts around the world. This information is analysed to identify EXE files associated with exploits of known vulnerabilities.

The extracted data is compared with public information about vulnerabilities which have been assigned CVEs. Researchers used this combination of information, along with other data sources such as AV telemetry data to correlate the date that vulnerabilities were disclosed to the date the corresponding exploits were seen in the wild, hence identifying zero-day attacks. Using these techniques, 18 zero-day vulnerabilities were found to have been exploited before disclosure. 11/18 of these zero-days were not known to have been used in attacks before until this research was performed.

Whilst these research papers provided techniques for zero-day detection and analysis, they offer no solution for preventing the sale of malware and zero-days on the DarkNet. Law enforcement cracking down on marketplaces and forums may assist in this effort. As more participants are arrested, others may be deterred from following the same path. However, current trends seem to suggest this isn't the case, there have been several marketplace takedowns and arrests but more are always ready to take their place, learning from the mistakes of their predecessors so that they don't fall for the same trap. A stronger effort from law enforcement is likely to drive markets deeper underground, rather than diminishing their popularity.

Bug bounties are a potential prevention method in terms of the zero-day market. If hackers feel they can make enough money legally by selling vulnerabilities directly to the affected vendor or to legitimate third-parties that deal in zero-days, this could help curb the success of black markets. Currently, one of the issues with bug bounties is that they are far less profitable than the black market, where traders could earn 10-100 times what the software vendor would offer. There can be large profits in the legal zero-day market though, one security researcher called "the Grugq" made around $1 million in 2012, acting as middleman between those with zero-days for sale and government agencies [9].

## VI. PREDICTIONS

Predictions indicate that the malware and zero-day markets on the DarkNet will continue to grow in popularity, particularly with respect to the as-a-service offerings which are proving to be profitable and are opening up the market to non-technical users. As media attention continues to increase, more users will be attracted, two factors which are likely to generate more attention from law enforcement and result in future takedowns and arrests. These actions will lead to the increased vetting of vendors and users and a stronger focus on encrypting and protecting communications and transactions.

The popularity of crypto-currencies is expected to continue rising, with new anonymous currencies emerging and those that are already established gaining further traction. As the current leaders of the DarkNet markets and forums retire, shift away from crime or get arrested, their places will be taken by digital natives who will have a greater understanding of the market and provide new levels of innovation. The presence of zero-days for sale is also expected to increase, as attacks increase in sophistication and more users flock to the markets creating an increased demand. However, bug bounties and zero-day detection methods could play a part in slowing the growth of this market area.

There are general cyber-security predictions which will have an effect on DarkNet markets. The population of the internet is increasing and new technologies such as Cloud, Mobile and IoT are rapidly developing, creating new attack vectors for hackers and malware authors. Traditional forms of data are gradually moving online as are traditional crimes e.g. fraud, narcotic sales and theft which increasingly have a digital component. Whilst many of these predictions are widely agreed upon in the security industry, others are disputed.

Some argue that hackers will target smaller organisations who are less likely to implement strong defences whilst others think they will go after larger targets where the rewards are greater. Some say that attacks seeking intellectual property will rise as the market is already flooded with digital accounts, identities and payment details. Some feel that low hanging fruit will dry up as the use of unsupported systems like Windows XP decreases but others say that the ever-increasing number of platforms and devices on the market will counteract this. Experts generally seem to agree that offence will continue to outpace defence.

## VII. CONCLUSION

DarkNet markets have become the primary medium for the distribution of malware and zero-days. Due to the anonymous nature of the DarkNet and the ability to buy products with anonymous crypto-currencies, this is unlikely to change in the near future. As these markets develop a larger customer base, hackers will continue to innovate. This phenomenon is having an impact on the overall security of the internet and currently law enforcement and security vendors are struggling to find practical solutions. Some ideas for detection and prevention have been mentioned in this paper and it is likely more will follow but as they do, cyber-criminals will adapt and evolve.